\documentclass[pdflatex,sn-mathphys-num]{sn-jnl}
\usepackage{amsmath, amssymb, graphics, epsfig, enumerate, amstext, amsthm, comment}
\usepackage{braket}
\usepackage{latexsym,bm,mathrsfs}
\usepackage{booktabs,makecell,array}
\def\diag{{\rm diag}}

\usepackage[dvipsnames]{xcolor}

\def\diag{{\rm diag}}

\newcommand{\Not}[1]{#1\kern-0.6em/}

\begin{document} 
\openup .8 \jot

\title
{An edge-based and subspace reduction encoding scheme to solve the traveling salesman problem in quantum computers }
\author*[1]{\fnm{Anandu} \sur{Kalleri Madhu}}\email{anandu.k.madhu@gmail.com}

\author[2]{\fnm{Chi-Kwong} \sur{Li}}\email{ckli@math.wm.edu}

\author[3]{\fnm{Jami} \sur{Rönkkö}}\email{jami@meetiqm.com}

\author[3]{\fnm{Mikio} \sur{Nakahara}}\email{mikio.nakahara@meetiqm.com}

\author[1,4,5,6]{\fnm{Ray-Kuang} \sur{Lee}}\email{rklee@ee.nthu.edu.tw}

\affil*[1]{\orgdiv{Department of Physics}, \orgname{National Tsing Hua University}, \orgaddress{ \city{Hsinchu}, \postcode{30013}, \country{Taiwan}}}

\affil[2]{\orgdiv{Department of Mathematics}, \orgname{College of William \& Mary}, \orgaddress{ \city{Williamsburg}, \state{VA 23187}, \country{USA}}}

\affil[3]{ \orgname {IQM Quantum Computers}, \orgaddress{ \street{Keilaranta 19}, \city{Espoo}, \postcode{02150}, \country{Finland}}}

\affil[4]{\orgdiv{Institute of Photonics Technologies}, \orgname{National Tsing Hua University}, \orgaddress{ \city{Hsinchu},  \postcode{30013}, \country{Taiwan}}}

\affil[5]{\orgdiv{Center for Theory and Computation}, \orgname{National Tsing Hua University}, \orgaddress{ \city{Hsinchu},  \postcode{30013}, \country{Taiwan}}}

\affil[6]{\orgdiv{Center for Quantum Science and Technology},  \orgaddress{ \city{Hsinchu},  \postcode{30013}, \country{Taiwan}}}

\date{}
\maketitle

\begin{abstract}

This paper introduces a novel edge-based encoding technique for solving the Traveling Salesman Problem (TSP) on a quantum computer, reducing the required number of qubits.  For implementation in real quantum devices, we applied the subspace reduction encoding to further reduce the dimension of the TSP solution space. We attack the TSP for 4-, 5-, and 6-city instances in both simulators and real quantum computers across different encoding frameworks. Optimal solutions of the 4-city TSP instance are obtained on state-of-the art IQM quantum computer. Our study presents a comparative analysis between edge-based encoding scheme and the node-based encoding methodology in the literature. Our findings indicate that the proposed encoding scheme outperforms conventional methods in terms of statistical measures, quantum resource utilization, and computational efficiency when applied to smaller TSP instances.  
\end{abstract}

\section{Introduction}

Quantum computing is rapidly emerging as a cutting-edge approach to solving optimization problems \cite{Bentley2022, Buonaiuto2023, Abbas2024, Silva2023}. Traveling Salesman Problem (TSP) is a combinatorial optimization problem that belongs to the NP-hard class. Hence, there does not exist a polynomial-time classical algorithm that can exactly solve this problem, unless P=NP. The generalized traveling salesman problem and vehicle routing problem \cite{Laporte1992a} are some generalizations of TSP. Some of the applications of TSP in real life include areas such as scheduling, logistics, transportation, DNA sequencing, and microchip design \cite{Matai2010, Kumar2003}. There are heuristic classical algorithms that give an approximate solution to TSP \cite{Laporte1992b, Malandraki1996}. But there are no practical classical algorithms that can provide an exact solution because the number of cities increases factorially with the problem size. Therefore, there is a huge scope for quantum algorithms to provide an improvement over current classical algorithms in terms of running time and accuracy of the solutions obtained. 

The common way to solve the TSP on a quantum computer is to represent the problem in quadratic unconstrained binary optimization (QUBO) format and map it into an Ising Hamiltonian-based quantum system like a quantum annealer \cite{Harris2018,PerdomoOrtiz2012,RolandCerf2002,Santoro2002}. There has also been considerable interest in simulating the TSP in gate-based quantum computers. Some of the quantum algorithms used for this are quantum phase estimation, Grover adaptive search, Variational quantum eigensolver (VQE) and Quantum approximation optimization algorithm (QAOA) \cite{Zhu2022, Khumalo2025}. Although these methods require considerable qubit overhead due to polynomial scaling with city count, recent investigations have explored quantum TSP algorithms with reduced qubit number requirements \cite{Srinivasan2018, Sato2025}, including approaches that employ only a single qubit \cite{Goswami2024}.

Recent work \cite{Salehi2022} has demonstrated the advantages of both edge-based and node-based formulations using quantum annealing for the TSP, inspired by the approach in \cite{papalitsas2019qubo}. Their formulations are based on a variation of the TSP that incorporates time-window constraints, known as the Traveling Salesman Problem with Time Windows (TSPTW). In this paper, we present two encoding schemes for representing a general TSP on a quantum computer, which we refer to as edge-encoding and subspace reduction encoding (SRE). The edge-encoding scheme is similar to the formulation in \cite{Salehi2022} in the sense that we encode the edges of the travel. However, our approach differs in how we construct the cost Hamiltonian. We use tensor products of diagonal matrices to build the cost Hamiltonian, encoding the travel costs directly as diagonal entries, which provides an easier construction by making use of diagonal vectors. We further apply both encoding schemes to simulate TSP instances on gate-based quantum computers.

For simulating the TSP on a quantum computer, we use a variational algorithm to find the minimum value of this cost Hamiltonian that represents the optimal travel of the TSP. We also compare our edge-based encoding scheme and SRE scheme with the node-based scheme, which represents the TSP as a QUBO or Ising Hamiltonian. We show that our edge-based scheme outperforms the node-based scheme for smaller TSP instances in terms of various statistical and computational metrics. The paper is organized as follows. In the Background section, we present the node-based encoding scheme as well as the QAOA variational algorithm. In the Methodology section, we present our edge-based scheme and the Subspace Reduction Encoding (SRE) scheme. In the Results section, we summarize all the results we obtained and discuss the comparison between the performance of the proposed encoding schemes and the conventional one. Finally, we summarize our findings and future directions in the conclusion.

\section{Background}

\subsection{Traveling Salesman Problem : Node-based encoding} \label{TSP statement}
The traveling salesman problem can be stated as follows.
Suppose a salesman visits $n$ cities starting in a city $i$ and
eventually returns to the same city $i$. His travel must
satisfy two constraints : (i) He visits all the cities once. (ii) He cannot visit the same city again.

Let $L_{ij}$ be the distance between the cities $i$ and $j$. For the travel schedule $i_0 \to i_1 \to \ldots \to i_{n-1}$, the total distance the salesperson travels is
\begin{equation}
L=\sum_{k=0}^{n-1} L_{i_k i_{k+1}}.
\end{equation} 
As mentioned above, the schedule satisfies the periodic boundary condition $i_{n}=i_0$. 

Let
\begin{equation}
A=\begin{pmatrix}
1&0\\
0&0
\end{pmatrix}
\end{equation}
be a matrix that has eigenvalues 1 and 0. The corresponding
eigenvectors are
\begin{equation}
\ket{0} =\begin{pmatrix}
1\\
0
\end{pmatrix},\quad 
\ket{1} = \begin{pmatrix}
0\\
1
\end{pmatrix}
.
\end{equation} 

Let
\begin{equation}
A_k = I_2 \otimes I_2 \otimes \ldots \otimes A \otimes I_2 
\ldots I_2,
\end{equation}
where $A$ is in the $k$th position and $I_2$ is the identity matrix of dimension 2.
The vector $\ket{0}$ corresponds to a city where the salesman stays while $\ket{1}$ corresponds to a city where he does not stay. The corresponding eigenvalues of $A$ are 1 and 0, respectively. Therefore, the salesman stays in the $k$th city if the eigenvalue of $A_k$ is 1 while not if the eigenvalue is 0.

\begin{figure}
    \centering
    \includegraphics[scale=0.45]{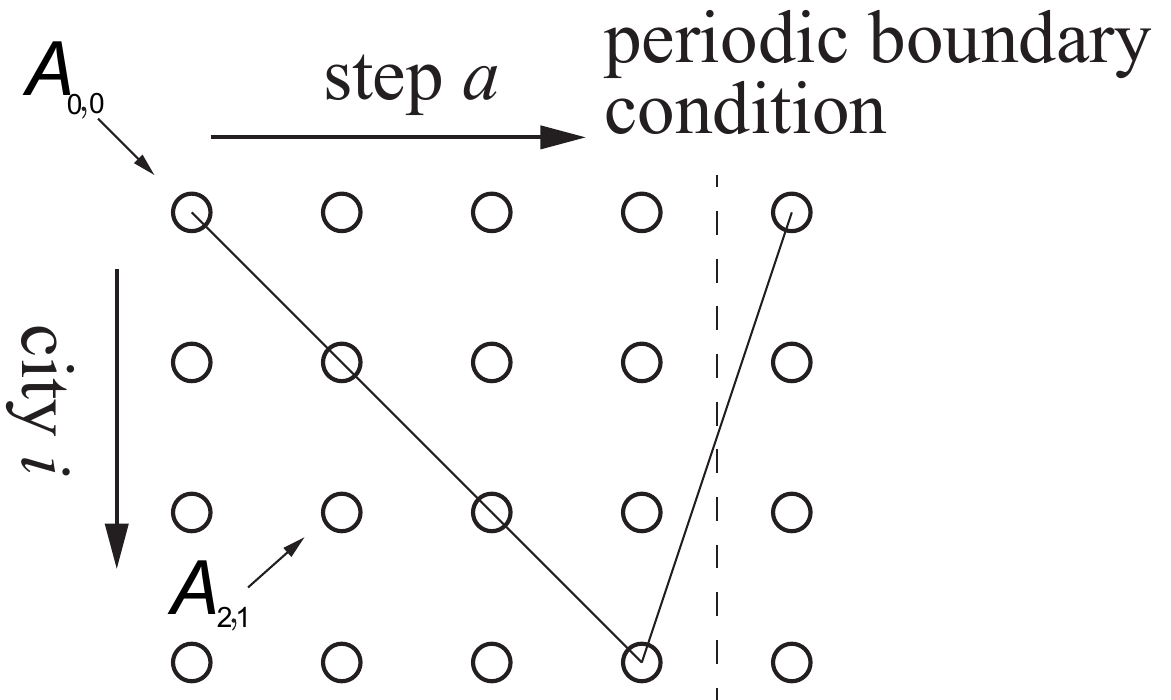}
    \caption{Alignment of cities for $n=4$.
    $\circ$ denotes a city and $i$ is the number of a city.
    $a$ denotes the step. For example, $A_{2,1}$
    denotes the 2nd city in the first step of the travel. The
    lines denote the flights the salesman takes. 
    This example corresponds to the travel $1 \to 2 \to 3 \to 4 \to 1$.
    }
    \label{fig:QUBOmap}
\end{figure}

Let $i_0 i_2 \ldots i_{n-1}$ be the path that the salesperson takes, where
it is understood that $i_{n}=i_0$. The distance of his travel is
$L=\sum_{k=0}^{n-1} L_{i_k i_{k+1}}$ as mentioned above.
Let us synthesize a Hamiltonian whose eigenvalues correspond to
the distances of travel.
For this purpose, we consider a column of $n$ cities and align
$n$ copies of this column to form a $n \times n$ matrix as
depicted in Fig. \ref{fig:QUBOmap}.
The fifth column is given in Fig. \ref{fig:QUBOmap} to make the periodic boundary
condition manifest.

Let us consider a cost Hamiltonian
\begin{equation}
H_\mathrm{node}= \sum_{i,a=0}^{n-1} L_{ij} A_{i,a} A_{j,a+1}, 
\end{equation}
where $L_{ij}=L_{ji}$ is the distance between the cities $i$ and $j$ and the index $a$ denotes the step of travel. 

We need to impose the two conditions;\\
(a) Each city is visited once and only once.
\\
(b) The salesman stays at one city in each step.
\\
We impose penalty for illegal travels that do not satisfy the above restrictions. This can be
done by adding extra terms to push up the eigenvalues
corresponding to them. Consider a modified cost Hamiltonian
\begin{equation}
\tilde H_\mathrm{node} = \sum_{i,a=0}^{n-1}L_{ij} A_{i,a} A_{j,a+1}
+\gamma \left[\sum_{a=0}^{n-1} \left( \sum_{i=0}^{n-1} A_{i,a}-I\right)^2
+\sum_{i=0}^{n-1} \left( \sum_{a=0}^{n-1} A_{i,a}-I \right)^2\right],
\end{equation}
where $\gamma$ is the penalty term and $I$ is the unit matrix of dimension $2^{n^2}$. 
For legal travels $\sum_{i=0}^{n-1} A_{i,a}$ and $\sum_{a=0}^{n-1} A_{i,a}$
have eigenvalues 1 and the penalty term vanishes although it is strictly positive for illegal travels. Therefore, if $\gamma \in \mathbb{R}$ is large enough, all the eigenvalues corresponding to illegal travels are pushed up high above those of legal travels. In effect, we can work in a subspace corresponding to legal travels, where $\tilde H_0$ is identified with $H_0$, and forget
illegal travels.

For a given travel schedule, there are $n$ choices of the first
city to start with and $2$ choices of the sense of travel, clockwise
or anticlockwise.

To encode a $n$-city TSP problem using this encoding in a quantum system, we will need to use the $n^2$ qubit system.

\subsection{Quantum Approximate Optimization Algorithm (QAOA)}

The Quantum Approximate Optimization Algorithm (QAOA), introduced by Farhi {\it et al.}~\cite{Farhi2014}, is a variational quantum algorithm constructed based on the theorem of adiabatic quantum computing. QAOA belongs to the class of hybrid quantum-classical algorithms, where a parameterized quantum circuit is used to prepare candidate solutions, and a classical optimizer is employed to tune the parameters to minimize a cost function.

QAOA seeks to minimize a classical cost function \( C(z) \) on bitstrings \( z \in \{0,1\}^n \). This objective function is encoded in a diagonal cost Hamiltonian \( H_C \), whose eigenvalues correspond to the cost of each bitstring. A mixing Hamiltonian \( H_M \), typically chosen as

\begin{equation}
H_M = \sum_{j=0}^{n-1} X_j,
\end{equation}
where \( X_j \) denotes the Pauli-X operator acting on the qubit \( j \), used to induce transitions between computational basis states. The QAOA variational ansatz at depth \( p \) is defined as

\begin{equation}
\ket{\psi(\boldsymbol{\gamma}, \boldsymbol{\beta})} = \prod_{j=1}^{p} e^{-i \beta_j H_M} e^{-i \gamma_j H_C} \ket{s},
\end{equation}
where $\ket{s} = \ket{+}^{\otimes n} $ is the initial uniform superposition state over all bitstrings, and \( (\boldsymbol{\gamma}, \boldsymbol{\beta}) \in \mathbb{R}^p \times \mathbb{R}^p \) are the variational parameters. The expectation value of the cost function is given by

\begin{equation}
C_p(\boldsymbol{\gamma}, \boldsymbol{\beta}) = \bra{\psi(\boldsymbol{\gamma}, \boldsymbol{\beta})} H_C \ket{\psi(\boldsymbol{\gamma}, \boldsymbol{\beta})},
\end{equation}
which is estimated via repeated quantum measurements. A classical optimizer is then used to iteratively update \( (\boldsymbol{\gamma}, \boldsymbol{\beta}) \) to minimize \( C_p \), guiding the quantum circuit to the optimal solution.

We use the QAOA algorithm to find the minimum eigenvalue of the TSP cost Hamiltonian which corresponds to the optimal travel of the TSP.
 
\section{Methodology}

In this section, we define the methodology of two encoding schemes that we propose to solve the TSP on a quantum system.

\subsection{Traveling Salesman Problem : Edge-based encoding} \label{sec : edge-based}

In this encoding scheme, the edges of the travel are used to construct the TSP Hamiltonian.
Suppose $n$ cities: $0, \dots, n-1$, are given. 
We define the TSP problem as a tour $i_0 - i_1 - i_2 - \cdots - i_{n-1}$ 
where $i_i$'s are cities such that $i_0, \dots, i_{n-1}$ are different and the schedule satisfies the  periodic boundary condition $i_0 = i_{n}$.

We always let $L = (\ell_{ij}) \in M_n$ as the cost matrix where $\ell_{ij}$ is the cost from city 
$i$ to  city $j$ if $i\ne j$ and $\ell_{jj} = 0$. Here, $M_n \in \mathbb{R}^{n \times n}$ denotes a real matrix of dimension \( n \times n \).
The key idea of our approach is to construct a diagonal matrix $H_\mathrm{edge}$ which is also the cost Hamiltonian so that the diagonal entries will record the cost of a collection of $n-1$ edges $(i_0,i_1), (i_1, i_2), \dots, (i_{n-2},i_{n-1})$
with severe penalty if these edges do not form a cycle with distinct vertices.
It is typically assumed that $L$ is symmetric, i.e., the traveling cost from city $i$ to city $j$ is the same as the cost from city $j$ to city $i$. However, there are real life problems, where the traveling costs between two cities are not symmetric. Our scheme does not assume that $L$ is symmetric.
 
\subsubsection{General Scheme} Given the travel schedule for $n$ cities as $(i_0,i_1,\dots, i_{n-1})$  
and the cost matrix $L = (\ell_{ij})$. 

\begin{enumerate}
\item 
Let $k$ be the smallest integer such that $n-1  \le 2^k$ and denote $2^k = K$. Here we start with city $0$ as the starting and ending city of the trip. We need to add some fake cities to the original set of cities in order to construct a matrix whose dimension is of the order $2^k$ or $K$. For this, we add $K-(n+1)$ fake cities and construct a $K\times K$  matrix $\tilde L = (\tilde \ell_{ij}) $ such that $\tilde \ell_{ij} = \ell_{ij}$ for $0 \le i, j \le n-1$, $\tilde\ell_{ij} = d$ if $i$ or $j$ is a newly added fake city, and then set $\ell_{jj} = d$ for all $j$, where $d$ is a large number used as a penalty to prevent the optimal tour from including a fake city, staying in a city twice or more, or not visiting a city at all.

\item Construct the cost Hamiltonian $H_\mathrm{edge}$ of size $K^{n-1} \times K^{n-1}$ which represents the cost of the 
schedules of the traveling salesman starting and ending the tour
with city 0, and going through cities $i_1,\dots, i_{n-1}$ as follows:

\begin{equation}
    H_\mathrm{edge} = D_1 \otimes I_K \otimes \underbrace{
I_{K}\otimes \cdots \otimes I_{K}}_{n-3}+  \sum_{j=1}^{n-1} C_j +\underbrace{
I_{K}\otimes \cdots \otimes I_{K}}_{n-3} \otimes  I_K  \otimes  D_2 + \gamma P,
\label{costH}
\end{equation}

where 
\begin{itemize}
\item[(1)] $D_1 = \diag(\ell_{01}, \dots, \ell_{0,n-1}) \oplus d I_{2^k-n} \in M_K$, 
 $D_2 = \diag(\ell_{10}, \dots, \ell_{n-1,0}) \oplus d I_{2^k-n} \in M_K$, 
 \item[(2)] $C_j$ is a tensor product of $C$ and $n-3$ copies of $I_{K}$ such that
$C_j$ appears as the $j$th component in the tensor product and 
$C \in M_{K^2}$ is a diagonal matrix with columns and rows indexed by
$r \in R =  \{(0,0), (0,1), \dots, (K,K)\}$ so that
$c_{rr} = \tilde\ell_{ij}$ if $r = (i,j)$.
\item[(3)]
$\gamma$ is a penalty 
constant (say, larger than $n$ times the maximum of $\ell_{ij}$ in the original
cost matrix), and $P\in M_{K^2}$ is a diagonal matrix with diagonal equal to 1 
if and only if it is indexed by $(r_1, \dots, r_{n-1})  \in R =  \{(0,0), (0,1), \dots, (K,K)\}$  with 

\medskip\centerline{ $r_{i} = r_{j}$ for some $0\le i<j \le K$.}
\end{itemize}

One can check that $D_1 \otimes I_K \otimes \underbrace{
I_{K}\otimes \cdots \otimes I_{K}}_{n-3}$ 
and $\underbrace{
I_{K}\otimes \cdots \otimes I_{K}}_{n-3} \otimes  I_K  \otimes  D_2$
will add the cost of the initial step $(0,i_1)$  and the final step $(i_{n-1},0)$ of 
the schedule $(0, i_1, \dots, i_{n-1})$.
The matrix $C_j$ will add the cost of the travel if $(i_j, i_{j+1})$ appears at
the schedule $(i_0,i_1,\dots, i_{n-1})$. Finally, the matrix $\gamma P$ will add the penalty 
to the tour if $i_r = i_s$ for some $1\le r<s \le K$ in the schedule 
$(i_1, \dots,  i_{n-1})$.

\item Then we can find the minimal eigenvalue (diagonal entry) of $H_\mathrm{edge}$ by QAOA. 

The ground state will occur at a diagonal entries labeled by
$(p_0, p_1, \dots, p_K)$ such that

\medskip\centerline{$p_0, p_1, \dots, p_K\in \{0, 1,\dots, K\}$ are different.}

\end{enumerate}

The cost matrix $H_\mathrm{edge}\in M_{K^{n-1}}$ with rows and columns indexed by 
$(i_1,\dots, i_{n-1})$ with $i_1, \dots, i_{n-1} \in \{0, \dots, K\}$. 
In implementation, each $i_r$ will be represented as a binary
number with $k$ digits, where $k = \lceil \log_2(n-1) \rceil$.
To encode an $n$-city TSP problem using this encoding scheme, we will need to use 
an $(n-1)\lceil \log_2(n-1) \rceil$ qubit system. 
Since this Hamiltonian contains many-body interaction terms, ideally a quantum device 
with all-to-all qubit connectivity will be suitable for simulating this Hamiltonian. 
Nevertheless, we can also use a quantum device with only local interactions to simulate 
this Hamiltonian using SWAP gates.

Evidently, we are able to reduce the number of qubits that are required to encode 
the TSP problem logarithmically as compared to the node-based scheme.

Also, it is easy to construct the cost matrix $H_\mathrm{edge}$ 
using tensor product of diagonal matrices (that are just the tensor product of
the vectors of the diagonal entries of matrices) so that one 
can use other algorithms to determine the ground state(s) once $H_\mathrm{edge}$ is set up.

\subsubsection{Example Problem : 4-city TSP}
\label{sec : 4c example}
Suppose  4 cities $0,1,2,3$ are given with the cost matrix
$L = (\ell_{ij}) = \begin{pmatrix}
0 & \ell_{01} & \ell_{02} & \ell_{03} \\
\ell_{10} & 0 & \ell_{12} & \ell_{13} \\
\ell_{20} & \ell_{21} & 0 & \ell_{23} \\
\ell_{30} & \ell_{31} & \ell_{32} & 0 \\

\end{pmatrix}$, which may or may not be symmetric.
We may assume that the travel starts at city 0 and ends at city 0.

\begin{itemize}
\item[1.] 
Let 
\begin{equation}
\label{gamma}
d > \max\{\ell_{ij}: 0 \le i,j \le 3\}
\quad \hbox{ and } \quad \gamma = 4d.
\end{equation}
Note that $n=4$ cities are given, and $n-1 < 2^2$.
Since the technique involves adding $(K - n + 1 )$ fake cities and here $K = 2^2$, we construct a fake city $\bar{4}$ (different from the real city 4), and construct
the new cost matrix from the original one:
$$\tilde L= \begin{pmatrix} 
d & \ell_{12} &  \ell_{13} & d\\
 \ell_{21} & d &  \ell_{23} & d \\
 \ell_{31} & \ell_{32} & d & d \\
 d & d & d & d \\
\end{pmatrix}.$$

Construct the diagonal matrix
$C \in M_{16}$ such that
the diagonal entry labeled by
$(i,j)$ with $(i,j) \in (1,1), (1,2), \dots, (4,4))$
 equals $\ell_{ij}$, the $(i,j)$ entry of $\tilde L$.
\item[2.]
\begin{equation} \label{C}
C = \diag(d, \ell_{12}, \ell_{13}, d, \ell_{21}, d, \ell_{23}, d,
 \ell_{31}, \ell_{32}, d, d, d, d, d, d).
 \end{equation}
Note that 10 diagonal entries of $C$ will equal to $d$ as $\hat L$ has 10 entries equal to $d$.
\item[3.]
Let 
\begin{equation}\label{D}
D_1 = \diag(\ell_{01}, \ell_{02}, \ell_{03}, d) , 
\quad D_2 = \diag(\ell_{10}, \ell_{20}, \ell_{30}, d)
\end{equation}
and
\begin{equation} \label{H}
H_\mathrm{edge} = D_1 \otimes I_{4} \otimes I_{4} + C \otimes I_{4} + I_{4} \otimes C +
I_{4} \otimes I_4 \otimes D_2 + \gamma P,
\end{equation}
where
 $D_1$ and $D_2$
will add the cost of the initial step $(0,i_1)$  and the final step $(i_3,0)$ of 
the schedule $(0, i_1, i_2, i_3)$ and
$P\in M_{64}$
is the diagonal zero one matrix with entry 0 at the diagonal positions 
 corresponding to the three cities where the traveling salesman stayed 
at the second, third, and the fourth time steps, i.e., 

$$(1,2,3), \ (1,3,2), \ 
(2,1,3),  \ (2,3,1), \
(3,1,2), \ (3,2,1). $$

Since a n-city TSP with the first city fixed contains $(n-1)!$ legal travels, here we have $3!=6$ legal travels or non-zero entries for the $P$ matrix.
Suppose the diagonal entries of $P$ is labeled by $0, \dots, 63$, or by
$(000000)$ to $(111111)$ in binary form. Then the 6 diagonal entries of $P$ equal to 0 will be
$$(000110), (001001), (010010), (011000), (100001), (100100).$$

Here, $P \in M_{64}$
and the diagonal entries are labeled by 
$(p_1,p_2,p_3) \in \{(1,1), (1,2), \dots, (4,4)\}$.
We can regard $(0,p_1,p_2,p_3)_4$ as a base 4 numbers so that 
$(0,0,0,0)_4 = 0$ and $(0,3,3,3)_4 = 63$.
then
the entry labeled by $(1,2,3)$ is $(0123)_4= 4^2+4*2+3 = 27$, the $27$th diagonal
entry. So, the above list of 6 diagonal entries of $P$ equal to 0 will be
$$ 27, 30, 39, 45, 54, 57
  $$

\end{itemize}

To encode this 4-city TSP example problem into a quantum computer we will require 6 qubits.

\subsection{Traveling Salesman Problem : Subspace Reduction Encoding (SRE)} \label{sec : SR Encoding}

Using the edge-based encoding scheme that we introduced, we were able to reduce the number of qubits that are needed to simulate the TSP problem logarithmically. But to further decrease the number of qubits required and the simulation time especially on the real devices, we had to reduce the dimension of the subspace that we are working with. If we know that the optimal solution of the problem lies in a certain subspace, it would be nice to find a scheme to improve the efficiency of a search algorithm to the subspace. For this, we developed a technique where we pre-select the legal travels of the TSP problem and constrain ourselves to search for the optimal travel within this subspace. This way, the optimizer of our variational algorithm could converge faster to the minimum value of the cost function and thereby find the optimal solution faster.

The algorithm is designed as follows :
\begin{itemize}
    \item[1.] For a $n$-city TSP problem, we have n! legal travels possible. In order to represent these $n!$ legal entries, we need a $n! \le 2^k = K$ dimensional  diagonal matrix. i.e., we require $k$ qubits to simulate this problem. 
    \item [2.] First, we pre-select the feasible TSP routes by generating all valid city permutations that correspond to legal tours. This can be accomplished using the \texttt{Permutations} function from the Python \texttt{Itertools} module, whose computational complexity scales as $\mathcal{O}{(n!)}$. Next, for each valid tour, we identify the associated diagonal indices of the cost Hamiltonian $H_\mathrm{edge}$ defined in Eq.~(\ref{costH}). The procedure for determining these indices follows the method described at the end of Section ~\ref{sec : 4c example}.
    \item[3.] We then create a new cost Hamiltonian ($H_{SRE} \in M_{K}$) with these diagonal entries labeled as $H_{SRE}(jj) = l_j$ where $l_j$ are the n! legal travels. We then pad the rest of the diagonal elements of this matrix with K-n! penalty entries to complete the matrix.
    \item[4.] Now, we can encode this cost Hamiltonian in a quantum computer to find the minimum eigenvalue of this Hamiltonian which corresponds to the optimal solution to our TSP using QAOA.
\end{itemize}

To encode a n-city TSP problem using this encoding, we only require to use a $\lceil \log_2(n!) \rceil$ qubit system. The efficiency of this scheme decreases for large problem sizes of the TSP since the algorithm scales as $\mathcal{O}{(n!)}$ with problem size. Nevertheless, this encoding scheme remains practical for the small problem instances that can currently be simulated on real quantum hardware.

\begin{figure}[h]
    \centering
    \includegraphics[scale=0.5]{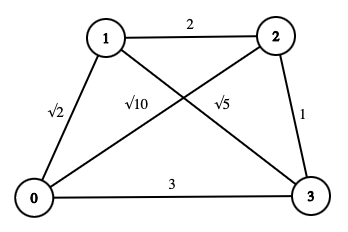}
    \caption{Four cities of the 4-city TSP and distances between pairs of cities.
$\circ$ denotes a city and the number attached to each line denotes the distance between two cities. }
    \label{fig:4c_TSP_layout}
\end{figure}

\begin{table}[h]
\centering
\begin{tabular}{
@{}%
>{\centering\arraybackslash}p{2.4cm}%
>{\centering\arraybackslash}p{2.4cm}%
>{\centering\arraybackslash}p{2.6cm}%
>{\centering\arraybackslash}p{1.4cm}%
>{\centering\arraybackslash}p{1.8cm}%
>{\centering\arraybackslash}p{2.6cm}@{}}
\toprule
\makecell{Number\\of cities} &
Encoding &
\makecell{Number\\of qubits} &
\makecell{SR} &
\makecell{Feas.\ \%} &
\makecell{Device\\Name} \\
\midrule
4 & node-based & 9  & 5  & 50 & simulator \\
4 & edge-based & 6  & 50 & 90 & simulator \\
\midrule
5 & node-based & 16 & 0  & 35 & simulator \\
5 & edge-based & 8  & 43 & 80 & simulator \\
\bottomrule
\end{tabular}
\caption{Comparison of QAOA performance on 4- and 5-city TSP instances using node-based and edge-based schemes. SR indicates the percentage of times the optimal route was found, and Feas. \% indicates the percentage of feasible solutions. Device name indicates the quantum device used for the simulation.}
\label{tab:qaoa_table}
\end{table}

\begin{figure}[h]
    \centering
    \includegraphics[scale=0.55]{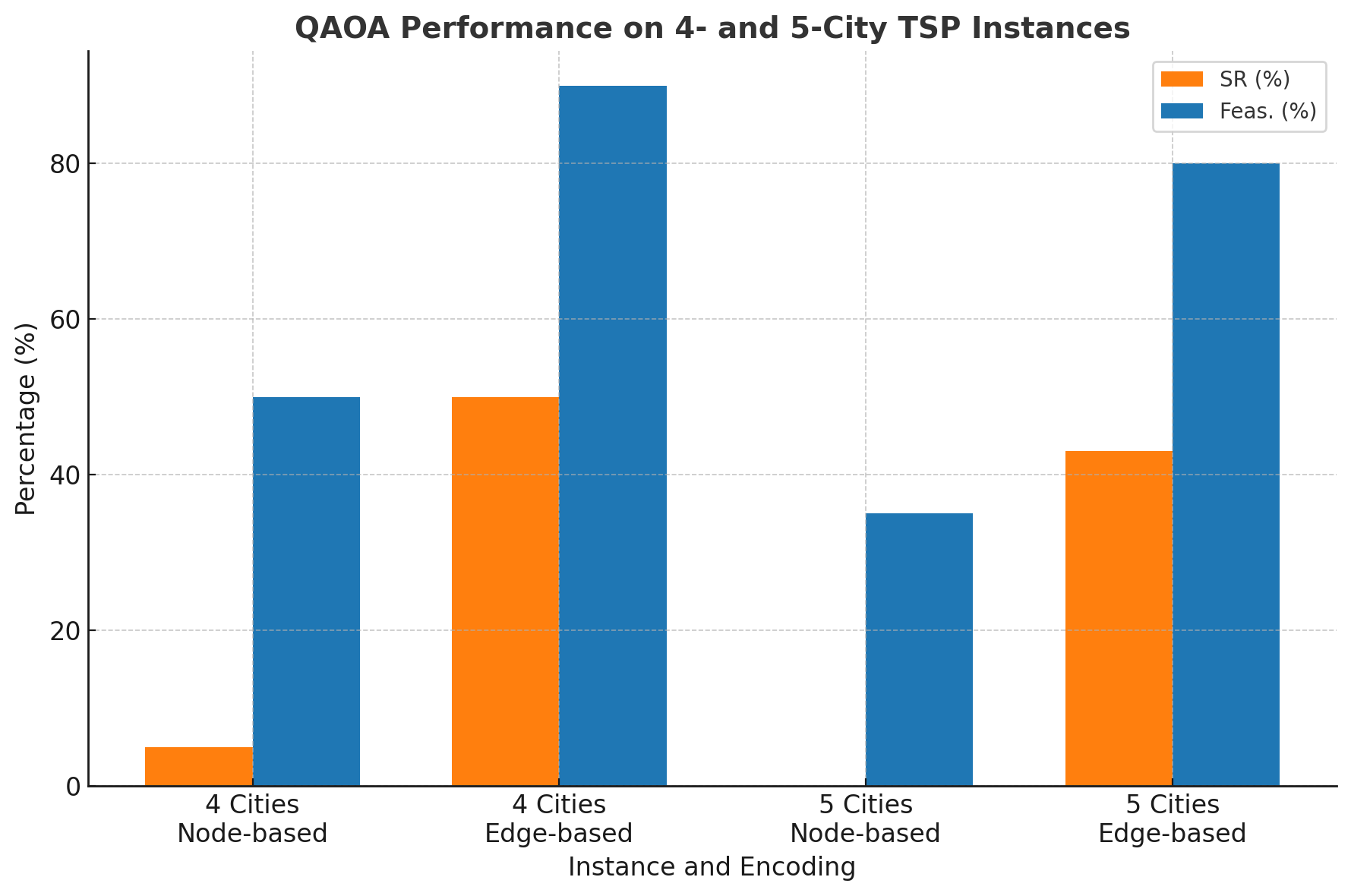}
    \caption{Comparison of QAOA performance on 4- and 5-city TSP instances using node-based and edge-based encoding schemes. The figure presents the Success Rate (SR) representing the percentage of times the optimal route was found, and the Feasibility percentage (Feas. \%) indicating the proportion of valid solutions obtained. All simulations were executed on the simulator.}
    \label{fig:qaoa_res_bar_plot}
\end{figure}

\section{Results}
 
To quantify the quality of the solutions obtained, we use different performance metrics to evaluate the results. We use Success Rate (SR) and Feasibility percentage (Feas. \%) to evaluate our results and compare the performance between different schemes \cite{Zhu2022}. The SR metric describes the percentage of trials that are within the optimal solutions of the TSP problem. Feasibility percentage calculates the percentage of trials that gives a feasible solution to the TSP problem. While implementing the QAOA algorithm in the quantum device, we specify the number of times (shots) the algorithm is sampled in the device. Therefore upon measurement, we obtain a distribution of states with different probabilities of occurrence. For a particular trial of the TSP algorithm, the most probable quantum state in the distribution is taken as the candidate solution.

 We encoded symmetric TSP instances of 4- and 5-city in simulators using our edge-based encoding scheme mentioned in Section \ref{sec : edge-based}. Fig. \ref{fig:4c_TSP_layout} denotes the layout and the distance between cities of the 4-city TSP. In both instances of the TSP, the starting city for travel is fixed. We use the QAOA algorithm to find the minimum value of the cost Hamiltonian. The results are given below in Table \ref{tab:qaoa_table} and in Fig. \ref{fig:qaoa_res_bar_plot}. One can see that the edge-based scheme requires less number of qubits for the simulation of TSP instances compared to the node-based scheme. Also, it performs better than the node-based scheme by the metrics SR and Feasibility percentage. We can see that the 4-city instance performs much better than the 5-city instance as expected since it uses less number of qubits and has less number of parameters in the variational algorithm to optimize. It has a Feasibility percentage of 90\% using the edge-based scheme. It also has a success rate of 50\% which means that half of the solutions obtained are optimal solutions. 

\begin{figure}[h]
    \centering
    \includegraphics[scale=0.5]{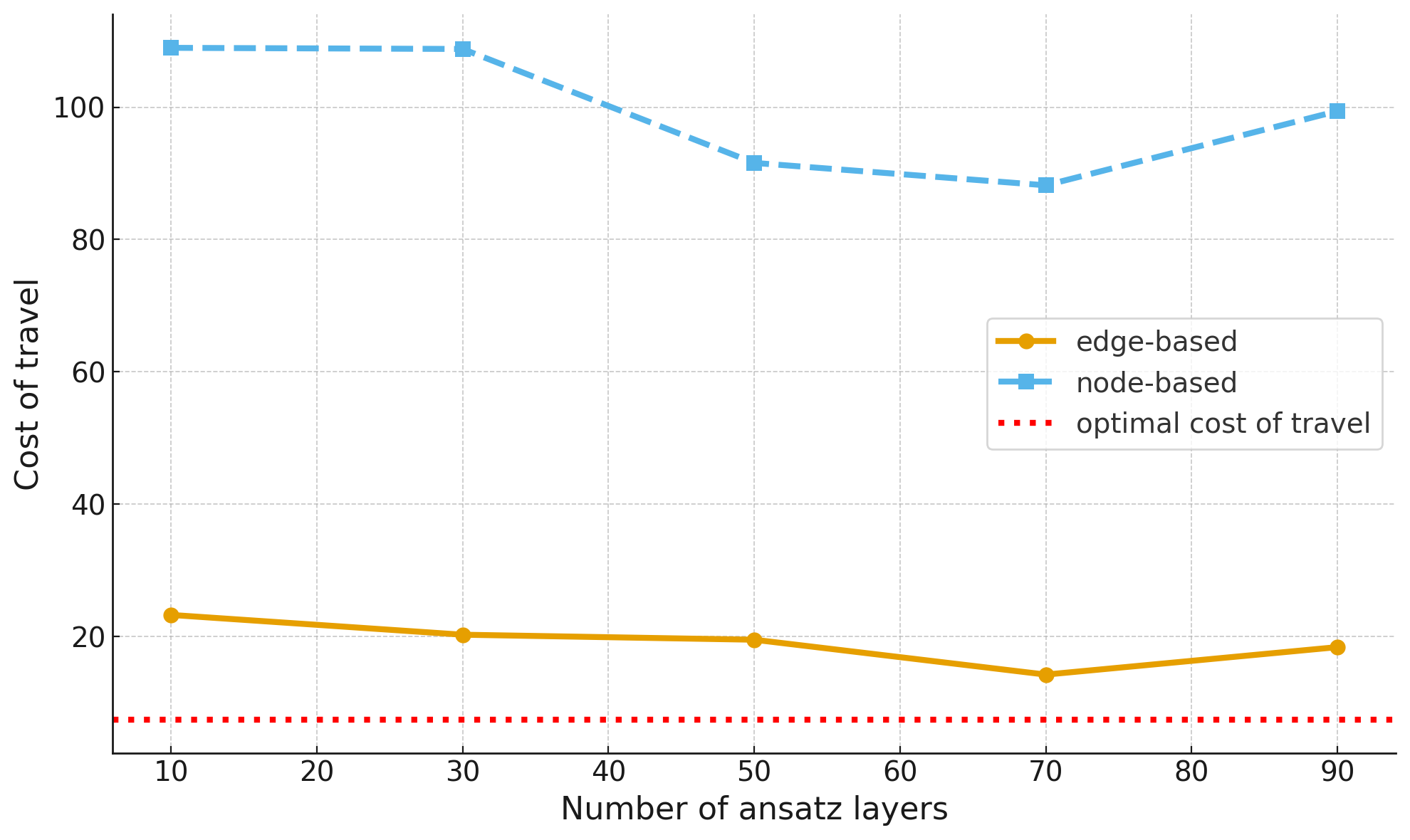}
    \caption{ Cost of travel as a function of the number of QAOA ansatz layers for the 4-city TSP, comparing two encoding schemes: edge-based and node-based.}
    \label{fig:4c_Energy_plot}
\end{figure}

\begin{figure}[h]
    \centering
    \includegraphics[scale=0.5]{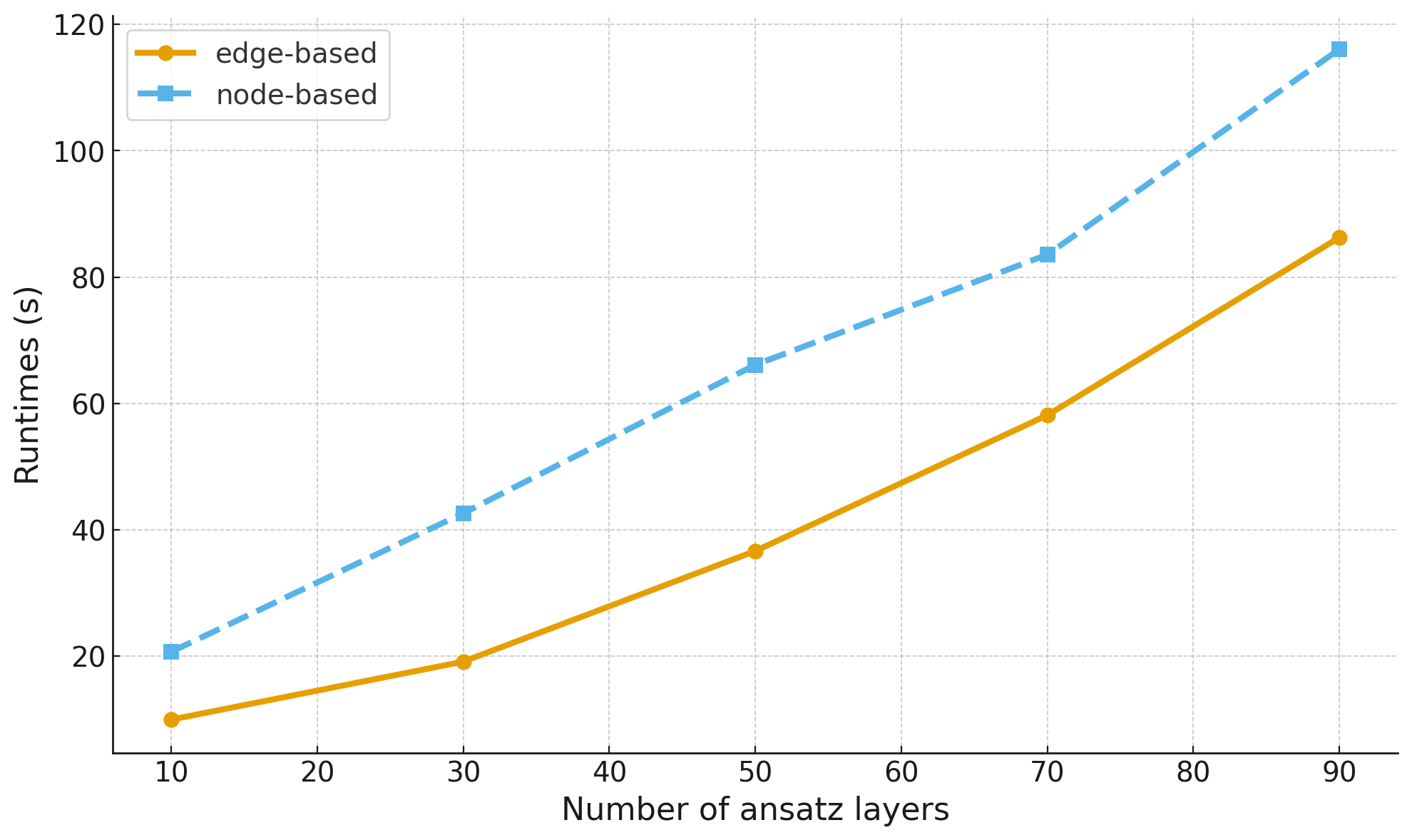}
    \caption{ Figure: Runtimes of the QAOA algorithm as a function of the number of QAOA ansatz layers for the 4-city Traveling Salesman Problem (TSP), comparing two encoding schemes: edge-based and node-based.}
    \label{fig:4c_runtime_plot}
\end{figure}

Fig. \ref{fig:4c_Energy_plot} and Fig. \ref{fig:4c_runtime_plot} shows the performance of the QAOA algorithm for different number of ansatz layers on the 4-city TSP instance. Fig. \ref{fig:4c_Energy_plot} shows the graph between the cost of travel and the number of layers of QAOA ansatz for both encoding schemes.  It is clearly noticeable that the optimizer of the QAOA algorithm is able to reduce the cost much better using our edge-based scheme as compared to the node-based scheme. Fig. \ref{fig:4c_runtime_plot} shows the graph between the runtime(s) of the algorithm and the number of layers of QAOA ansatz for both encoding schemes. We can see that it takes much less time for the algorithm to converge to the optimal solution for the edge-based scheme as compared to the node-based scheme. Therefore our edge-based scheme performs better in terms of reducing the energy or cost in a lesser amount of time. This improvement can be attributed to the fact that the edge-based scheme restricts the problem to a smaller subspace, requiring fewer qubits.

\begin{figure}[h]
    \centering
    \includegraphics[scale=0.55]{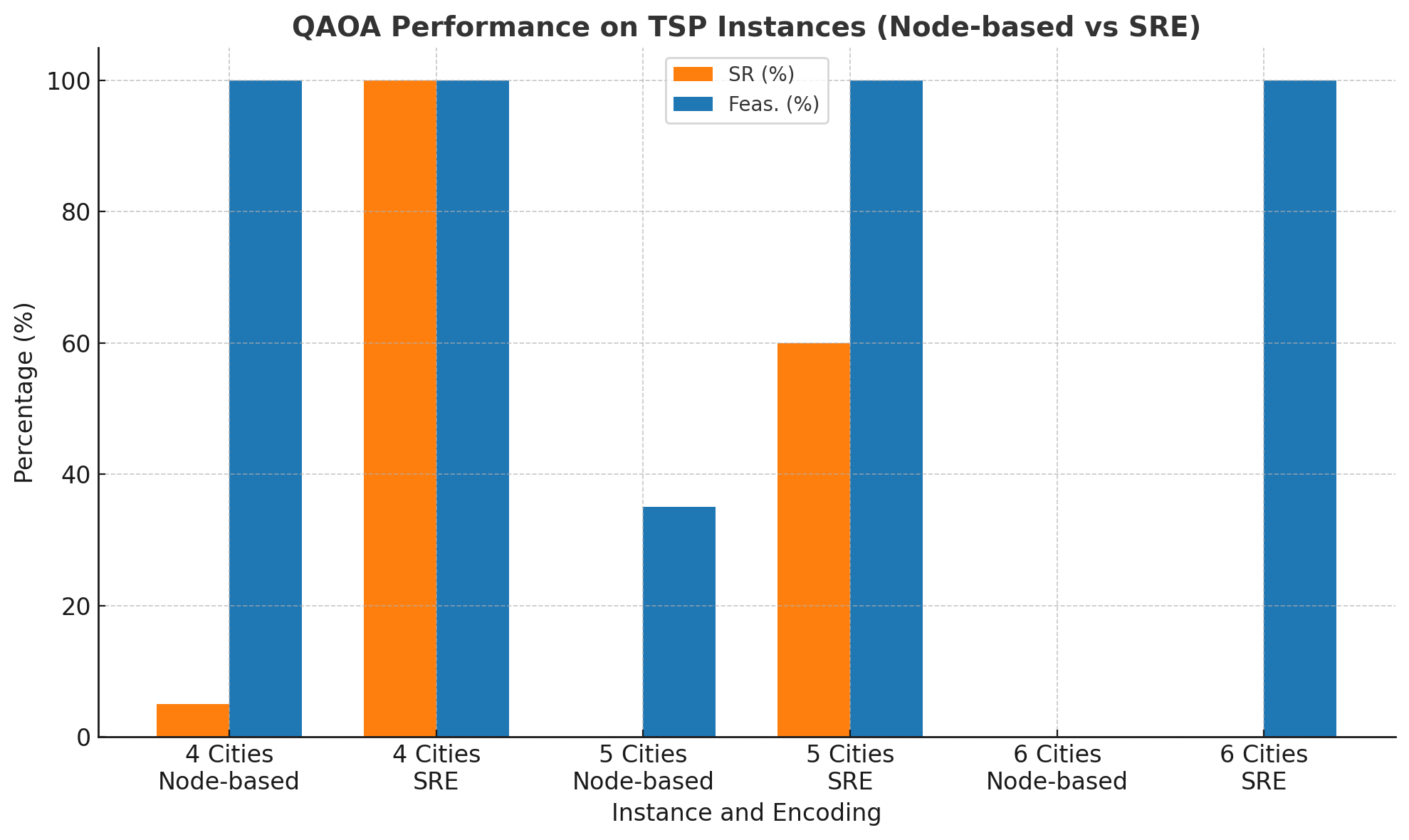}
    \caption{Comparison of QAOA performance on 4-, 5-, and 6-city TSP instances using node-based and SRE encoding schemes. The figure presents the Success Rate (SR) representing the percentage of times the optimal route was found, and the Feasibility percentage (Feas. \%) indicating the proportion of valid solutions obtained. All simulations were executed on the Aer simulator.}
    \label{fig:qaoa_res_sre_bar_plot}
\end{figure}

In order to reduce the simulation time further, we used the Subspace Reduction Encoding mentioned in Section \ref{sec : SR Encoding} to simulate symmetric TSP instances of 4-, 5- and 6-city. The first city of travel has been fixed in this case as well. The results of these implementations on quantum simulators are shown in Table. \ref{tab:oe_sre_table} and Fig. \ref{fig:qaoa_res_sre_bar_plot}. We can see that the Subspace Reduction Encoding scheme uses much less number of qubits as compared to the node-based scheme. This reduced subspace implies better results according to the metrics SR and Feasibility Percentage as shown in Table. \ref{tab:oe_sre_table}. The SR for 4-city TSP instance is 100\% using SRE scheme. We also have improvement in the simulation results for the 5-city TSP instance as compared to the edge-based and node-based schemes. Also, for the first time we were able to simulate the 6-city TSP using this encoding. Although we couldn't find any optimal solution for this instance, we obtained 100\% feasible solutions, which is a big improvement compared to the node-based scheme. Using SRE scheme, we also could reduce the runtime of the algorithm by a considerable amount. 

The main reason to introduce the SRE scheme was to implement the TSP problems efficiently in real quantum computers. We were able to successfully implement the 4-city TSP instance in state-of-the-art IQM quantum computer, IQM Garnet, and obtained optimal solution as shown in Table. \ref{tab:qaoa_sre_table}. To save resources, we first implemented the TSP instances on simulators using QAOA and find optimal parameters of the ansatz. We then simulated the QAOA ansatz on the real quantum computers using these optimal parameters to sample our solutions. We implemented the QAOA with a single layer of ansatz. This is because more than a single layer of ansatz is shown to give inconsistent results due to large gate errors in the real device quantum circuits. Due to this limitation, we could not find an optimal solution to the 5- and 6- city TSP problem in the real devices.

\renewcommand{\arraystretch}{1.2}

\begin{table}[h]
\centering
\begin{tabular}{
@{}%
>{\centering\arraybackslash}p{2.0cm}%
>{\centering\arraybackslash}p{2.4cm}%
>{\centering\arraybackslash}p{2.4cm}%
>{\centering\arraybackslash}p{1.8cm}%
>{\centering\arraybackslash}p{2.0cm}%
>{\centering\arraybackslash}p{2.6cm}@{}}
\toprule
\makecell{Number\\of cities} &
Encoding &
\makecell{Number\\of qubits} &
\makecell{SR\\} &
\makecell{Feas.\ \%} &
\makecell{Device\\Name} \\
\midrule
4 & node-based & 9  & 5   & 100 & simulator \\
4 & SRE        & 3  & 100 & 100 & simulator \\
\midrule
5 & node-based & 16 & 0   & 35  & simulator \\
5 & SRE        & 5  & 60  & 100 & simulator \\
\midrule
6 & node-based & 25 & 0   & 0   & simulator \\
6 & SRE        & 7  & 0   & 100 & simulator \\
\bottomrule
\end{tabular}
\caption{Comparison of QAOA performance on TSP instances using node-based and SRE schemes. SR indicates the percentage of times the optimal route was found, and Feas.\% indicates the percentage of feasible solutions. Device name indicates the quantum device used for the simulation.}
\label{tab:oe_sre_table}
\end{table}

\renewcommand{\arraystretch}{1.2}

\begin{table}[h]
\centering
\begin{tabular}{
@{}%
>{\centering\arraybackslash}p{2.0cm}%
>{\centering\arraybackslash}p{2.2cm}%
>{\centering\arraybackslash}p{2.4cm}%
>{\centering\arraybackslash}p{2.0cm}%
>{\centering\arraybackslash}p{2.6cm}@{}}
\toprule
\makecell{Number\\of cities} &
Encoding &
\makecell{Number\\of qubits} &
\makecell{Probability\\(\%)} &
\makecell{Device\\Name} \\
\midrule

4 & SRE & 3 & 33.58 & IQM Garnet \\
\midrule
5 & SRE & 5 & -- & IQM Garnet \\
\midrule
6 & SRE & 7 & -- & IQM Garnet \\
\bottomrule
\end{tabular}
\caption{QAOA performance on TSP instances using SRE scheme implemented on real quantum computer `IQM Garnet'. `Probability \%' indicates the probability of obtaining the optimal solution of the TSP for 10,000 shots on the real device. 5- and 6- city TSP simulations were not able to obtain the optimal solutions. Device name indicates the real quantum device used for the implementation of the TSP.}
\label{tab:qaoa_sre_table}
\end{table}

\section{Conclusion}

In this paper, we proposed two encoding schemes called edge-based encoding and subspace reduction encoding for solving the Traveling Salesman Problem and compared it with the existing node-based encoding scheme in the literature. Using the edge-based scheme, we were able to reduce the number of qubits that are required to simulate a TSP instance logarithmically compared to the node-based approach, and we were also able to simulate different instances of the TSP using different encoding schemes in quantum simulators and evaluate their performance. The results we obtained indicate that our proposed edge-based scheme performed much better than the node-based scheme in all the metrics that we evaluated, and we also found that our edge-based scheme also has lesser simulation time compared to the node-based scheme. We also proposed another encoding scheme called the Subspace Reduction Encoding (SRE) scheme which is useful for simulating small TSP instances in real quantum computers due to its fast simulation runtime and less usage of qubits. We were able to successfully implement and obtain optimal solutions to the 4-city TSP instance in real quantum computers. The encoding scheme that we proposed is useful for simulation of Traveling Salesman Problem with a small number of qubits, and this encoding strategy doesn’t assume the TSP to be symmetric. Our encoding technique saves important computational resources such as qubits and algorithm runtime. 
Yet, there are many rooms for improvements for the new techniques as complexity of finding an optimal solution for the TSP still increases exponentially with the problem size. From an experimental point of view, building a quantum device with all-to-all qubit connectivity will improve the efficiency of implementing our newly proposed schemes. Nevertheless, we hope that our research paves the way for creating more novel encoding strategies for the Traveling Salesman Problem and related combinatorial optimization problems.

\section{Acknowledgments}

C.-K.L would like to thank Dr. Nung-Sing Sze of the Hong Kong Polytechnic University for some helpful discussion. M.N. and J.R. would like to thank Juha Vartiainen for his encouragement and interest in this work.

\section{Declarations}

\subsection{Funding}
C.-K.L was partially supported by the Simons Foundation Grant 851334. R.-K.L was partially supported by the National Science and Technology Council of Taiwan (Nos 112-2123-M-007-001, 112-2119-M-008-007, 114-2112-M-007-044-MY3),  Office of Naval Research Global, the International Technology Center Indo-Pacific (ITC IPAC) and Army Research Office, under Contract No. FA5209-21-P-0158, and the collaborative research program of the Institute for Cosmic Ray Research (ICRR) at the University of Tokyo. 

\subsection{Conflict of interest}
The authors declare no competing interests.

\subsection{Ethics approval and consent to participate}
‘Not applicable’

\subsection{Consent for publication}
‘Not applicable’

\subsection{Data availability}
Data and code associated with this study has been deposited
at https://github.com/anandukmadhu/TSP

\subsection{Materials availability}
‘Not applicable’

\subsection{Code Availability}
Code associated with this study has been deposited
at https://github.com/anandukmadhu/TSP

\subsection{Author contribution}
Anandu Kalleri Madhu and Chi-Kwong Li conceived and designed the experiments. Anandu Kalleri Madhu and Jami Rönkkö performed the experiments. Anandu Kalleri Madhu, Chi-Kwong Li, Jami Rönkkö, Mikio Nakahara, and Ray-Kuang Lee analyzed and interpreted the data. All authors contributed to writing and revising the manuscript.
\bibliography{references}        

\end{document}